\documentstyle[preprint,aps]{revtex}
\tighten 
\begin{document}
\title{Effective Magnetic Moment of Neutrinos in Strong Magnetic Fields}
\author{Samina.S.Masood\thanks{samina@physics.sdnpk.undp.org}$^{1,2}$,
A. P\'erez-Mart\'{\i}nez\thanks{aurora@cidet.icmf.inf.cu}$^{2,3}$, 
H. Perez Rojas $^3$ R. Gait\'an$^4$ and S. Rodriguez-Romo $^4$ \\
1.International Center for Theoretical Physics ICTP, Trieste, Italy\\
2.Phys. Dept. Quaid-i-Azam University, Islamabad, Pakistan\\     
3.Instituto de Matem\'atica Cibern\'etica y F\'{\i}sica, calle E
esq. a 15 No. 309.Vedado C. Habana,Cuba\\
4.Centro de Investigaciones Te\'oricas, FES, UNAM,\\ A. Postal 142,
Cuautitl\'an-Izcalli, Estado de M\'exico, \\
C. Postal 54700, M\'exico.}
\maketitle
\begin{abstract}
We compute the effective magnetic moment of neutrino in the highly dense and strongly magnetized media.
It is shown that this magnetic moment is generated due to the effective mass of neutrino and gives sufficiently high value of the magnetic moment in the core of neutron stars.
\end{abstract}

\vspace{1 cm}

The study of the effects of intense magnetic fields on various 
astrophysical phenomenon has  got a considerable importance nowadays.
Large magnetic fields of the order $B_{m}\sim 10^{12}-10^{14}$ $Gauss$ is
being associated with the surfaces of supernovae \cite{sup} and neutron
stars  \cite{2}-\cite{3}.
It is also known that this strong magnetic field quantize the motion of 
electrons in the interior of neutron stars \cite{neu}. The dispersion
curves for
neutrinos propagating in such an extremely dense plasmas in the strong 
magnetic fields of order $ B\le m_{W}^2/e$ has already been calculated
\cite{yo}. 
By starting form the standard model neutrinos we have shown  that these
neutrinos while propagating through a very dense electroweak plasma in the
presence of a very strong magnetic field ($eB\le m_{W}^2$) acquire large
effective mass $m_{eff}$. These effectively massive neutrinos exhibit a
large effective magnetic moment while passing through this highly magnetized 
medium.
The effective magnetic dipole moment describes the neutrino energy shift in a magnetized medium as a consequence of the $\gamma$-structure of the self-energy which is not that of a magnetic dipole interaction 
\cite{rafel}.
We work in the framework of imaginary time formalism and replace the 
vacuum propagators with the corresponding background corrected  propagators
in presence of the strong magnetic field. All the Feynman rules of the
 vacuum theory remain the same, otherwise.
The electron propagator in the configuration space with the background corrections comes out to be \cite{hug}
\[
G^e(x,x^{\prime})=-\frac{1}{(2\pi^2)\beta}\sum_{n^{\prime}=0}^{\infty}%
\sum_{p_{4}} \int\frac{dp_{3}dp_{2}} {(p_{4}^{*^2}+p_{3}^2+m_{e}^2+2eBn^{%
\prime})} 
\]
\[
\cdot\left\{ \left( ip_4^{*}\gamma _4+ip_3\gamma _3-m_{e}) (\sigma _{+}\psi
_{n^{\prime}}(\xi)\psi_{n^{\prime}}(\xi^{\prime})+\sigma _{-}\psi
_{n^{\prime}-1}(\xi) \psi _{n^{\prime}-1}(\xi^{\prime}) \right) \right. 
\]
\begin{equation}
+1/2\sqrt{2eBn^{\prime}}[\gamma _{+}\psi _{n^{\prime}}(\xi)\psi
_{n^{\prime}-1}(\xi^{\prime})- \left. \gamma _{-}\psi
_{n^{\prime}-1}(\xi)\psi _{n^{\prime}}(\xi^{\prime}) ]\right\}  \label{ele}
\end{equation}
\[
\cdot{\rm exp}[ip_{4}(x_{4}-x^{\prime}_{4})+ip_{3}(x_{3}-x^{\prime}_{3})
+ip_{2}(x_{2}-x^{^{\prime}}_{2})], 
\]
\noindent
where $\xi=\sqrt{eB}(x_{1}+x_{o})$, $\xi^{\prime}=\sqrt{eB}(x_{1}^{\prime}
+x_{o})$, $x_{o}=\frac{p_{2}}{eB}$, $\sigma ^{\pm }=1/2[1\pm \sigma _3]$,
 $\gamma_{\pm}=1/2[\gamma _1\pm i\gamma _2]$,
 $\sigma _3 =i/2[\gamma _1,\gamma _2]$, 
$p_{\lambda}^{*}=p_{\lambda}-i\mu\delta_{4\lambda}$, $n^{\prime}$ is the
electron Landau quantum number and $\psi_{n^{\prime}}(\xi)$ are Hermite functions for  $A_{\mu}=(0,Bx,0,0)$.
The propagator of W-bosons in the magnetic field has the form
\begin{equation}
D_{\mu\nu}^W(x,x^{^{\prime}})=\frac{1}{(2\pi)^2\beta}\sum_{p_{4}}
\sum_{n} \int dp_{2}dp_{3} [\frac{R^{-}+R^{+}}{2}\Psi^{1}_{\mu\nu}+R^0%
\Psi^{2}_{\mu\nu}+ i\frac{(R^{-}-R^{+})}{2}\Psi^{3}_{\mu\nu})]  \label{dmu}
\end{equation}
\[
\cdot \psi_{n}(\xi)\psi_{n}(\xi^{\prime}){\rm exp}[ip_{4}
(x_{4}-x^{\prime}_{4}) +ip_{3}(x_{3}-x^{\prime}_{3})
+ip_{2}(x_{2}-x^{\prime}_{2})]. 
\]
\noindent
where $n$ is the $W$ Landau quantum number, $R^{\pm}= [p_{4}^{*^2} +
E_{n}^{W2} \pm 2eB]^{-1}$, $R^0=[p_{4}^{*2} + E_{n}^{W2}]^{-1}$, with
$E_{n}^{W2} = M_{W}^2 + p_{3}^2 +2eB(n+1/2)$; $\Psi^{1}_{\mu\nu}=\frac{1}{B^2}
G^{0 2}_{\mu\nu}$, $\Psi^{2}_{\mu\nu}=\delta_{\mu\nu}-\frac{1}{B^2} G^{0
2}_{\mu\nu}$ and $\Psi^{3}_{\mu\nu}=\frac{1}{B} G^{0}_{\mu\nu}$ ($G_{\mu\nu}^{02}$ is the field tensor of the $SU(2)\times U(1)$ electromagnetic
external field). Concerning the gauge fixing term, we are taking $D^W_{\mu\nu}$ in a transverse gauge which is expected to guarantee the gauge
independence of the neutrino spectrum.
The charged current contribution to the self-energy of neutrino is 

\begin{equation}
\Sigma^{C}(x,x^{\prime})=-i\frac{g^2}{(2\pi)^3}\gamma_{\mu}
G^{e}(x,x^{\prime})D^{W}_{\mu\nu}(x-x^{\prime})\gamma_{\nu},  \label{sig}
\end{equation}
\noindent
whereas  the effective mass of neutrino can be calculated from the
inverse propagator of  the neutrino in the momentum space given by,  \begin{equation}
\Sigma^{C}(k)=\frac{g^2eB}{(2\pi)^2}(\sum_{p_{4}}\int dp_{3}
G^{o}_{e}(p_3+k_3,p_4+k_4)\Sigma_{\alpha\beta})P_{L},  \label{mag}
\end{equation}
\noindent
where  $P_{L}$ is the usual left projection operator.
By solving
\begin{equation}
{\rm det}(-i\gamma_{\mu}k_{\mu}+\Sigma^{W})=0,  \label{dis}
\end{equation}

\noindent
the effective mass of neutrino $m_{eff}$ becomes a function of the
magnetic field itself in highly magnetic superdense medium \cite{yo},
giving

\begin{equation}
m_{eff} =\frac{g^2e|B|\mu_{e}}{(2\pi)^2(m_{W}^2-e|B|)}
\end{equation}

\noindent
where $\mu_{e}$ is the electron chemical potential which we assume very 
large ($\mu_{e}\approx 10^2m_{e}$).
The expression (6) is valid whenever $\sqrt{M^2_{W}-eB}>>\mu_{e}$.  
This Dirac type effective mass of neutrinos can couple in the medium 
with the external magnetic field through real electrons.
The corresponding value of the magnetic dipole moment of neutrino can 
then be evaluated as
\begin{equation}
d_{\nu}=\frac{d}{d}\frac{m_{eff}}{|B|}\label{ee}
\end{equation}

\noindent
which come out to be equal to

\begin{equation}
d_{\nu}=\frac{g^2 e\mu_{e}m_{W}^2}{(2\pi)^2 (m_{W}^2-e|B|)^2}\label{dip1}
\end{equation}
\noindent
giving  $d_{\nu}=0.79\times 10^{-10} \mu_{B}$ for
 $B_{m}\sim 10^{12}-10^{14}$ $Gauss$ inside the neutron stars.
 Notice also that when the magnetic field (locally) rises to 
$0.9m_{W}/e\sim 10^{23}$ $Gauss$ the corresponding magnetic moment increases
to $0.79 \times 10^{-8} \mu_{B}$.
It is interesting to note in Eq.(\ref{dip1}) that the magnetic moment of neutrino  
becomes a function of the magnetic field due to the $m_{eff}$ dependence on 
$B$.
It is different from the dipole magnetic moment correction due to 
statistical background  \cite{sami} for massive neutrinos which always 
appear as a 
coefficient of the magnetic field.
It can also be seen from Eq.(\ref{dip1}) that if we  expand this equation 
for small values of $B<<m_{W}$, we  still get a nonzero  value of the effective dipole moment 
$d_{\nu}=g^2e\mu_{e}/m_{W}^2=0.79\times 10^{-10}\mu_{B}$ which is generated
due to the effectively massive neutrinos. 
It can also be seen from Fig.1 that the magnetic
moment of neutrinos at $B\sim 10^{12}-10^{14}$ $Gauss$  correspond to this constant value.

This low $B$ value of the magnetic moment will only be significant 
at large electron densities $\mu_{e}=2\pi^2N_{e}/eB$. 
This behaviour of $d_{\nu}$ as a function of $B$ has been plotted in Fig.1.
It can be seen from this plot that for very strong fields the magnetic 
moment changes rapidly with the field, so the dependence of  $d_{\nu}$ with regards to $B$ becomes stronger with the increasing magnetic field.

From Fig.1 and also directly from Eq.(\ref{dip1}) it seems  
that the magnetic dipole moment  of neutrino has a pole at 
$B\sim m_{W}^2/e \sim 10^{24}$ $Gauss$ which is the required primordial 
background field to galactic dynamo effect and is needed to produce 
electroweak phase transition on the
scale of one correlation length $1/m_{W}$ \cite{9}.

However this singularity is only apparent, since for $(M_{W}^2-eB)\approx \mu_{e}$, the approximation on which (6) is based, ceases to be valid. Also for  fields extremely near $B_{c}$ the production of pairs $W^{\pm}$
in the ground state is increased, a $W^{+}$ condensate  appears which 
modifies Eq.(\ref{dip1}) and as a result $m_{eff}$ does not actually 
diverge \cite{yo}. 
 
\vspace{1cm} 

\section{Acknowledgments}
Two of us (S.M and A.P.M) would like to acknowledge the partial support by
Abdus Salam ICTP, A.P.M, H.P.R and R. Gaitan thanks the support of CONACYT 
proyect and the Caribbean Network on Quantum Mechanics Particles and Fields
from OEA of ICTP.

\begin{figure}
\caption{Variation of the Effective Magnetic Dipole Moment as a function of B}
\label{fig1}
\end{figure}
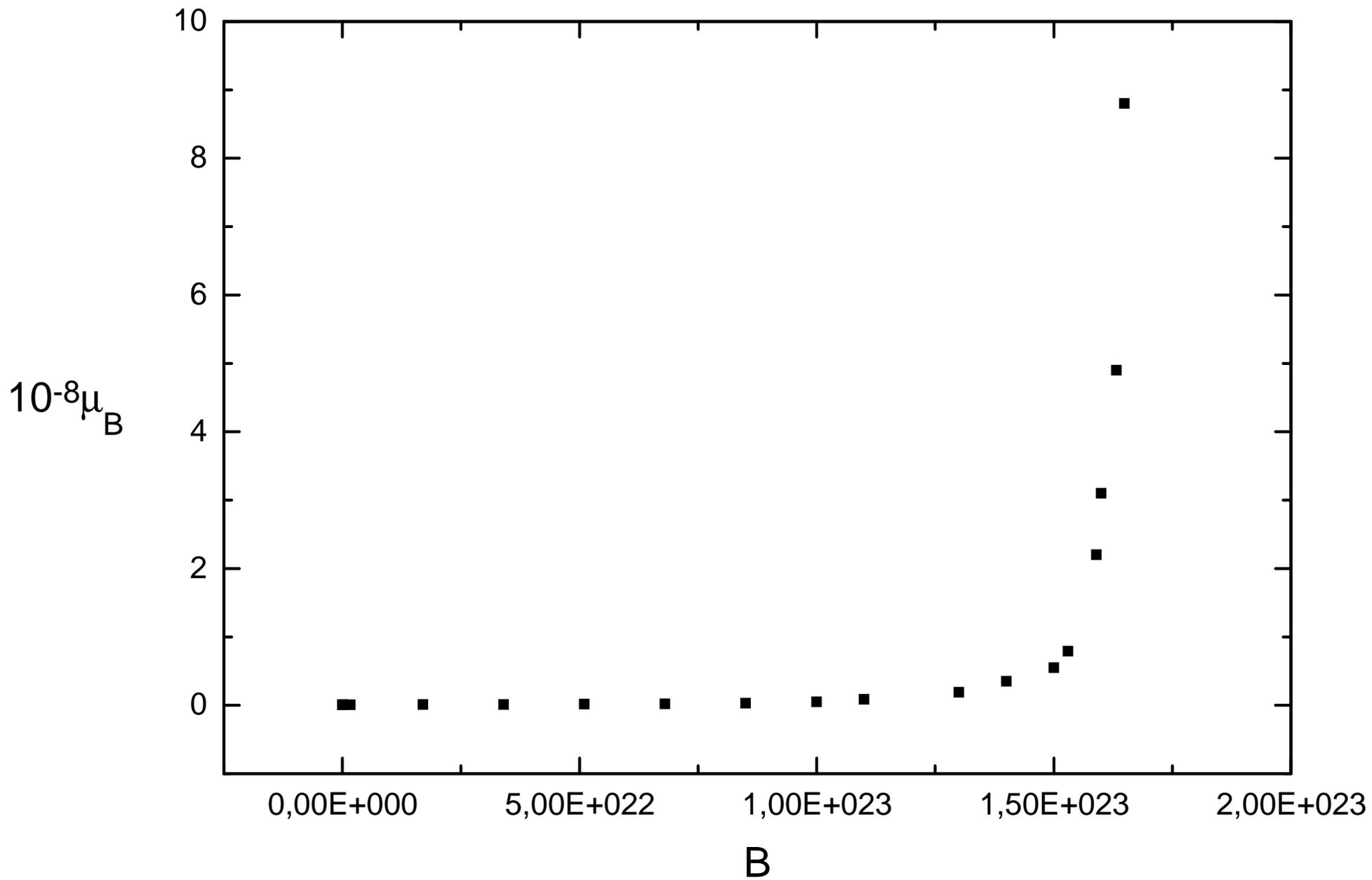
\end{document}